# Learning in Multi-Level Stochastic Games with Delayed Information

Edward A. Billard
Faculty of Computer Science and Engineering
University of Aizu
Aizu-Wakamatsu City, 965-80 Japan
E-mail: billard@u-aizu.ac.jp

## Abstract

Distributed decision makers are modeled as players in a game with two levels. High level decisions concern the game environment and determine the willingness of the players to form a coalition (or group). Low level decisions involve the actions to be implemented within the chosen environment. Coalition and action strategies are determined by probability distributions which are updated using learning automata schemes. The payoffs are also probabilistic and there is uncertainty in the state vector since information is delayed. The goal is to reach equilibrium in both levels of decision making; the results show the conditions for instability, based on the age of information.

## 1 Introduction

Agents in a distributed system make decisions to optimize a performance metric or achieve a more abstract set of goals. These agents must typically consider working with other agents to cooperatively achieve the desired result. However, there is a high degree of uncertainty in these activities. First, the agent may not know the true state of the system as a result of delayed information. The delays may be due to inherent latencies in a network or the intermittent (or periodic) exchange of information. The agents make the best possible decisions with the information available [Gmytrasiewicz et al., 1991c]. Second, even with instantaneous information, there is uncertainty in the strategies employed by the other agents given the state vector. For example, an agent may not be certain that another agent is willing to cooperate or to what extent. Third, even with knowledge of the other strategies, there is uncertainty in the payoffs that result from the combined actions.

We present a model to capture the nature of these various uncertainties with distributed decision makers as players in a game with two levels. The high level concerns the game environment and determines the willingness of the players to form a coalition (or group). The low level involves the actions to be implemented within the chosen environment.

Both of these strategies are modeled using probability distributions with updates according to learning automata schemes [Narendra and Thathachar, 1989]. This implies that learning is taking place on two levels and a constraint is that a player must make both decisions simultaneously, without knowledge of the other players' decisions at either level. In particular, a player knows whether it is willing to form a group but does not know the intentions of the other players. This implies that a player may select an action under the assumption of cooperative behavior but this action, in the context of non-cooperative behavior, may result in suboptimal performance.

The payoffs in the games are stochastic, that is, there is a probability of gain or loss based upon the action set. Uncertainty in information is captured by the assumption that an average age of information exists in the system. The goal of the model is to capture decision making under uncertainty in various domains and to summarize uncertainty as probability distributions. The adaptive learning schemes easily model the uncertainty, permit expected value computations to determine beliefs, and have analytic solutions to complex dynamical behaviors. These schemes may also be considered as approximations to more complex reasoning schemes.

In most distributed systems an important goal is to achieve a stable solution. We develop a dynamical equation to predict the behavior based on the parameter settings and apply linear stability analysis to predict the onset of persistent oscillations.

The paper is organized as follows: Section 2 describes related work; Section 3 develops the model in stages, including the dynamical equation; Section 4 shows example simulations and associated predicted behavior. In Section 5, we make an assumption that leads to a reasonably accurate prediction of the delay required to initiate persistent instabilities in the system. Our conclusions are presented in Section 6.

## 2 Related Work

Our interests in distributed decision making are closely related to the evolution of cooperation [Axelrod and Hamilton, 1981] and computational ecosystems. The original description of computational ecologies [Huberman and Hogg, 1988] shows the dynamical equation based on simple gain functions with imperfect and delayed information. A large system of agents select resources based on aged information of other agents' resource preferences. The resultant behavior can be categorized as stable, oscillatory (both damped and persistent), or chaotic (with possible bifurcations). The agreement between the dynamical equation and simulation is demonstrated in [Kephart et al., 1989] and the existence of a general adaptive strategy to eliminate the instabilities is shown in [Hogg and Huberman, 1991].

In distributed computing systems, a high degree of physical decentralization leads to aged information such that agents are not able to attain common knowledge [Halpern and Moses, 1990]. The goal of agents in these systems is to make good decisions with the information available and, in particular, to make good decisions involving cooperation with other agents. Other research examines cooperation without communication [Genesereth et al., 1985] and cooperation with negotiated protocols [Rosenschein and Genesereth, 1985].

Our approach is to examine learning mechanisms such as learning automata [Narendra and Thathachar, 1989] in environments with delayed information. The basic research relevant to automata playing stochastic games (and the associated dynamics) is found in [Lakshmivarahan and Narendra, 1982]. Our model extends this to delayed information and a hierarchy of games. The games in our model represent the payoffs of an underlying application such as robotics [Gmytrasiewicz et al., 1991a].

Learning automata have demonstrated coadaptive behavior in a distributed queueing system [Glockner and Pasquale, 1993]. We have also examined learning automata in autonomous decentralized queueing systems [Billard and Pasquale, 1993a] and in games [Billard and Pasquale, 1993b]. We view the learning algorithms as generic in the sense that they capture incremental, or adaptive, learning.

Although increased levels of communication can reduce the age of information to the minimum latency, there is an associated cost in processing this information. For this reason, it is important to exchange only the appropriate information. This can be done based on expected utility [Gmytrasiewicz et al., 1991c] with agents reaching equilibrium using recursive reasoning [Gmytrasiewicz et al., 1991b].

## 3 The Model

The model is developed in four stages: 1) the basic algorithm for a learning automaton [Narendra and Thathachar, 1989], 2) the algorithm applied to the strategies of two players in a game, 3) the algorithm applied again to the strategies of selecting between two games, and 4) the delay in state information. The salient feature of the model is that each agent makes a decision to work in a group or alone, thus affecting the environmental payoffs, and a decision regarding the action to be taken within the chosen environment.

*Step 1: One Automaton - Two Strategies*

Let $p(t)$ and $\bar{p}(t)$ be the probability of selecting strategy 1 and strategy 2, respectively, at time $t$. The probability is incremented or decremented for the next time step by

$$\Delta p = \theta \cdot \begin{cases} +\beta\bar{p} & \text{if reward on strategy 1} \\ -\beta p & \text{if reward on strategy 2} \\ -\alpha p & \text{if penalty on strategy 1} \\ +\alpha\bar{p} & \text{if penalty on strategy 2} \end{cases} \quad (1)$$

The extent of the incremental change in the mixed strategy is determined by the three constants: $\beta$ is the reward parameter, $\alpha$ is the penalty parameter, and $\theta$ is the step size parameter. It is assumed that $0 < \alpha < \beta < 1$ and $0 < \theta \leq 1$. Although $\theta$ can be incorporated into $\alpha$ and $\beta$, it is convenient to extract this term for simulation and analysis results.

*Step 2: Two Players - Two Strategies*

We define two players $k, l \in \{1, 2\}$ in a game $\mathbf{D} = (\mathbf{D}^1, \mathbf{D}^2)$, where $\mathbf{D}^k$ represents a stochastic payoff matrix for player $k$ [Lakshmivarahan and Narendra, 1982; Narendra and Thathachar, 1989] and corresponds to an underlying application. Each player chooses a strategy $i, j \in \{1, 2\}$, respectively, and the game is played in stages with element $d_{ij}^k$ of $\mathbf{D}^k$ being the probability of a unit gain for player $k$ based upon the strategy pair $(i, j)$. With probability $1-d_{ij}^k$, player $k$ receives a unit loss. This differs from games with deterministic payoffs as there is uncertainty in the result based upon the strategy pair. In the model, the game payoffs are the expected difference in gain and loss, $g_{ij}^k = 2d_{ij}^k - 1$, which scales to the interval $[-1, +1]$. The bi-matrix $\mathbf{D}$ is a nonzero-sum game such that both players may receive a unit gain (or unit loss), that is, $d_{ij}^1$ does not necessarily equal $1-d_{ij}^2$.

The decisions are made using randomization and, as such, both players are uncertain as to the pure strategy that will be employed by the other player. Let $\mathbf{p} = (p_1, p_2)$ be the state vector where $p_k$ is the probability that player $k$ will select strategy 1 and $\bar{p}_k$ is the probability of strategy 2. Each player employs an automaton to update the probabilities for the next stage where a unit gain is a reward and a unit loss is a penalty.

The following closely parallels the derivation in [Laksh-mivarahan and Narendra, 1982] except that we include nonzero-sum games, delayed information, and a more general notation that permits learning in a hierarchy of games.

Let $\delta z(t) = z(t+1) - z(t)$. The expected change in the probability vector can be deduced from (1). For example, with probability $p_1$, player 1 will select strategy 1. If the player receives a reward, then $p_1$ will increment by $\theta\beta\bar{p}_1$. Following this reasoning for all possibilities:

$$E[\delta\mathbf{p}(t)|\mathbf{p}(t) = \mathbf{p}] = \theta\mathbf{W}(\mathbf{p}), \qquad (2)$$

where

$$\begin{aligned} W_k(\mathbf{p}) &= \beta p_k \bar{p}_k [C_1^k(\mathbf{p}) - C_2^k(\mathbf{p})] + \\ &\quad \alpha[\bar{p}_k^2 \bar{C}_2^k(\mathbf{p}) - p_k^2 \bar{C}_1^k(\mathbf{p})] \end{aligned} \qquad (3)$$

and $C_i^k(\mathbf{p})$ is the probability that player $k$ receives a reward for strategy $i$. This is determined as follows. Let $\mathbf{p}_k = (p_k\ \bar{p}_k)$ be the probability vector for player $k$. The expected game payoff, or value of the game, for player $k$ is

$$\eta_k(\mathbf{p}) = \mathbf{p}_1 \mathbf{D}^k \mathbf{p}_2^T, \qquad (4)$$

where $\mathbf{p}_2^T$ is the transpose of $\mathbf{p}_2$. Now, $C_i^k(\mathbf{p}) = \eta_k(\mathbf{q})$ where $\mathbf{q} = \mathbf{p}$ but with the $k$th element replaced by $2$-$i$. For example, if player 1 selects strategy 1, then the expected payoff is $p_2 d_{11}^1 + \bar{p}_2 d_{12}^1$.

We recast the difference equation as a differential equation as this closely captures the behavior for the typical parameter settings, i.e. small $\theta$. Therefore,

$$\frac{d\mathbf{p}}{dt} = \theta\mathbf{W}(\mathbf{p}). \qquad (5)$$

The equilibrium solution is $\mathbf{p}^*$ where $\mathbf{W}(\mathbf{p}^*) = 0$. Note that the values of the learning parameters affect the equilibrium solution, that is, $\mathbf{p}^* = f(\alpha, \beta, \mathbf{D})$.

*Step 3: Four Players  -  Two Games*

We introduce the concept of multi-level games to capture the notion of cooperation in group dynamics, see Figure 1. An agent consists of two subcomponents, or players, each of which is modeled as a learning automaton. One player within each agent makes a preference decision between two game bi-matrices **A**, the *non-default* game matrix, and **B**, the *default* game matrix. Game **B** represents the underlying environment when the agents choose not to form a group. Typically, the payoffs will be lower but easier to achieve (in the sense of an equilibrium). Game **A** represents the environment when both agents agree to cooperate in a group with the expectation that better payoffs are available to both agents. However, to achieve these payoffs, the agents must successfully coordinate their actions within the game, perhaps a more difficult task in this game than in **B**. This second activity, i.e. selecting an action strategy within the chosen game environment, is carried out by an additional player within each agent. If an agent is willing to play game **A**, there is uncertainty whether the other agent will agree and, hence, the player subcomponent may make poor action decisions. For example, player 1 may select action strategy 1 since it has a high expectation of success in game **A**, the agent's preferred matrix. If agent 2 forces the default game environment, strategy 1 may yield a very poor result. It is the uncertainty in coalition formation and the simultaneity of decision making that makes action decision making difficult.

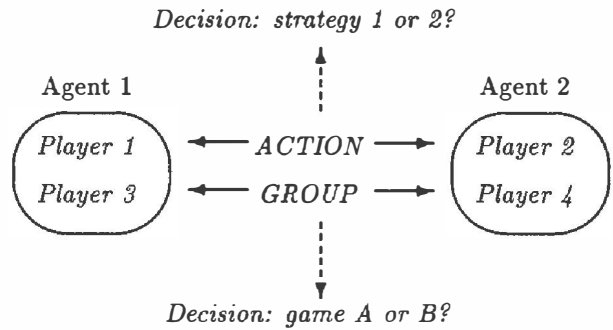

Figure 1: Multi-Level Decision Making by Agents' Components

We define the high level decisions (i.e. *which* game matrix) as group strategies and the low level decisions (i.e. which strategy *within* a game) as action strategies. The formal definition of the model is as follows.

The action strategies are determined as before (using $p_1$ and $p_2$). The group strategies are also made using randomization with $p_3$ the probability that player 3 (a subcomponent of agent 1) will prefer **A** over **B** (likewise, $p_4$ is the probability for player 4, a subcomponent of agent 2). The state vector is now $\mathbf{p} = (p_1, p_2, p_3, p_4)$. At the high level, each player uses an automaton to decide the game preference. At the low level, each player uses a different automaton to select a strategy. The action pair is determined at the same time as the group decision. The resultant action pair $(i, j)$ is played in game **A** if, and only if, *both* agents prefer this game matrix. That is, the agents agree to form a coalition with probability $c = p_3 p_4$, the clustering parameter. Otherwise, the stochastic payoffs are determined by **B** with the agents operating in a non-coalition mode. The problem of apportioning credit to the different levels is avoided by assuming that both levels receive the same payoff, that is, both receive either a unit gain or unit loss.

An average game is induced based on the high level strategies:

$$\mathbf{D}^k = c \cdot \mathbf{A}^k + (1 - c) \cdot \mathbf{B}^k. \qquad (6)$$

The dynamical equation is still (5) but where $k \in \{1, 2, 3, 4\}$ and $\mathbf{D}^k = \mathbf{D}^{k-2}$ for $k \in \{3, 4\}$. Note that this equation enforces a strong interaction among the state variables. The low level strategies are dependent on the high level strategies for the expectation of the average game. They are also dependent on each other via the stochastic payoffs based on action pairs. The high level strategies are dependent on the low level strategies since the reward (or penalty) is derived in the same way. The potential exists for different learning rates at different levels but, in this study, both rates are identically $\theta$.

|  $A_1$  |  | $B_1$ |  |  $A_2$  |  | $B_2$ |  |  $A_3$  |  | $B_2$ |  |
|---|---|---|---|---|---|---|---|---|---|---|---|
| 1,1 | 0,0 | 0,0 | 0,0 | .6,.4 | .2,.8 | .4,.6 | .8,.2 | .75,1 | .5,.25 | .4,.6 | .8,.2 |
| 0,0 | 0,0 | 0,0 | 1,1 | .35,.65 | .9,.1 | .65,.35 | .1,.9 | 1,.5 | .25,.75 | .65,.35 | .1,.9 |
| *game 1* |  |  |  | *game 2* |  |  |  | *game 3* |  |  |  |

Figure 2: Example Games

*Step 4: Delayed Information*

Since agents are physically distributed, the information available to an agent is delayed. The state vector **p** describes the probabilities of decisions at both the high and low level and, in our model, is subject to aged information. That is, the agents must make the best decisions possible given an aged view of the likelihood of the other agent's decisions.

Let $\tau$ be the average delay in information, representing the overall effect of latency within the distributed system. For example, latency is increased by periodic broadcasts of information or by the inherent delays within network hardware and software. The latency is a fundamental cause of uncertainty.

Consider a probability $p_k(t)$. We define an aged view of this probability as $p_k^\tau = p_k(t-\tau)$ where $p_k(t) = p_k(0)$ for $t < 0$. Agent $k$ knows with certainty the probability of its low and high strategies, $p_k$ and $p_{k+2}$, respectively, and has an aged view of the other two probabilities. From the subcomponents point of view, let $\mathbf{p}^k$ be player $k$'s view of the state vector, that is, $\mathbf{p}^1 = \mathbf{p}^3 = (p_1, p_2^\tau, p_3, p_4^\tau)$ and $\mathbf{p}^2 = \mathbf{p}^4 = (p_1^\tau, p_2, p_3^\tau, p_4)$.

In terms of the rules of the game, the preceding implies that a local module, or score keeper, provides a unit gain or loss based on the decisions of the local agent and the aged probabilities of the distant agent. For example, the local module for agent 1 determines the outcome based on the agent's pure strategy $i$ and chance, but where chance is now determined by $p_2^\tau d_{i1} + \bar{p}_2^\tau d_{i2}$ (and the average game element $d_{ij}$ is also based on aged information).

Now, (5) may be applied using $W_k(\mathbf{p}^k)$ instead of the instantaneous vector **p**. For example, the rate of change in $p_1(t)$ is a function of $p_1(t), p_3(t), p_2(t-\tau)$, and $p_4(t-\tau)$. Formally, (5) is a nonlinear delay differential equation [Wiener and Hale, 1992].

## 4  Experiments

Three games, see Figure 2, are considered with respect to learning behavior and the stability of the probabilistic strategies. The games are chosen to facilitate the illustration of key points and do not necessarily represent an underlying application. In game 1, the high level choice is between two game matrices, both with pure strategy equilibria of identical payoffs to both players. However, an opposite set of actions is required to achieve equilibrium. In game 2, the matrices are complements of each other and both are zero-sum game matrices with mixed strategy equilibria (the single game is from [Lakshmivarahan and Narendra, 1982]). In game 3, one choice is a nonzero-sum game matrix with mixed strategy equilibrium and the other is the same default game matrix of game 2.

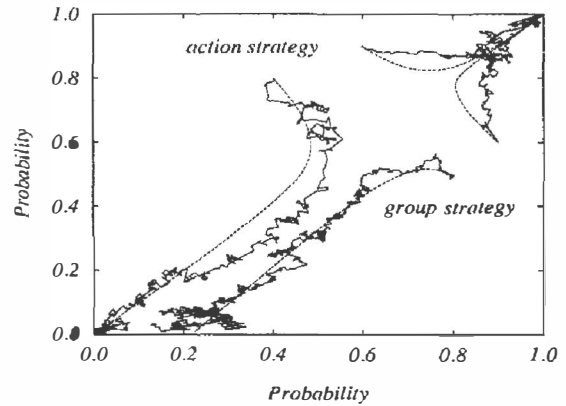

Figure 3: Phase-Plane Portrait of Game 1 with Two Pure Strategy Equilbria ($\tau = 0$)

Figure 3 shows the action and group strategies for game 1 in two experiments with different initialization (the delay in the system is zero.) The action strategies are plotted as $p_2$ versus $p_1$ and the group strategies as $p_4$ versus $p_3$. The initialization determines which of the two pure equilibria is "closest". The single runs roughly approximate the predicted behavior based on a numerical solution to (5), that is, the players are able to reach an equilibrium in both levels. Note that the group strategy for the non-coalition equilibrium does not terminate at the origin. Instead, both strategies decrease at the same linear rate and whichever strategy reaches zero first (based on initialization) prevents coalition formation (i.e. $c=p_3p_4=0$). This linear behavior is due to the contrived nature of the game payoffs.

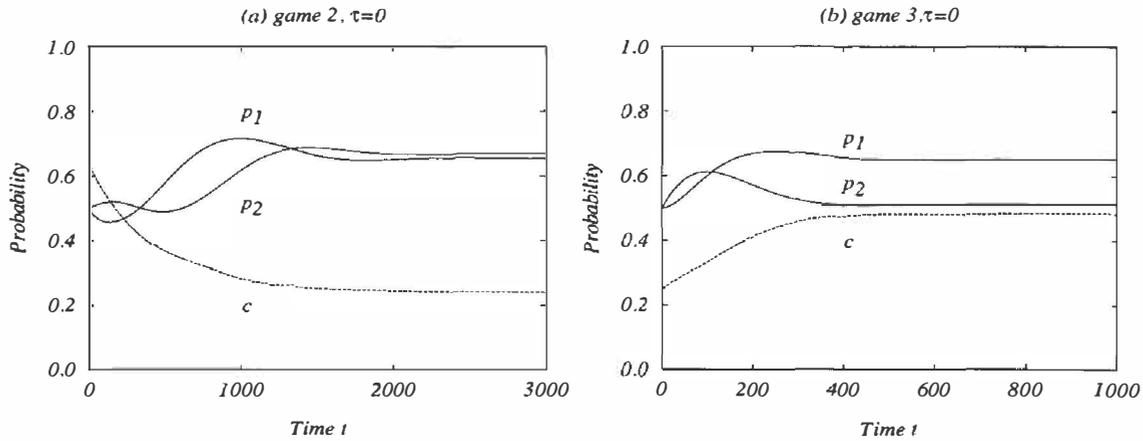

Figure 4: Equilibria in Action and Group Mixed Strategies

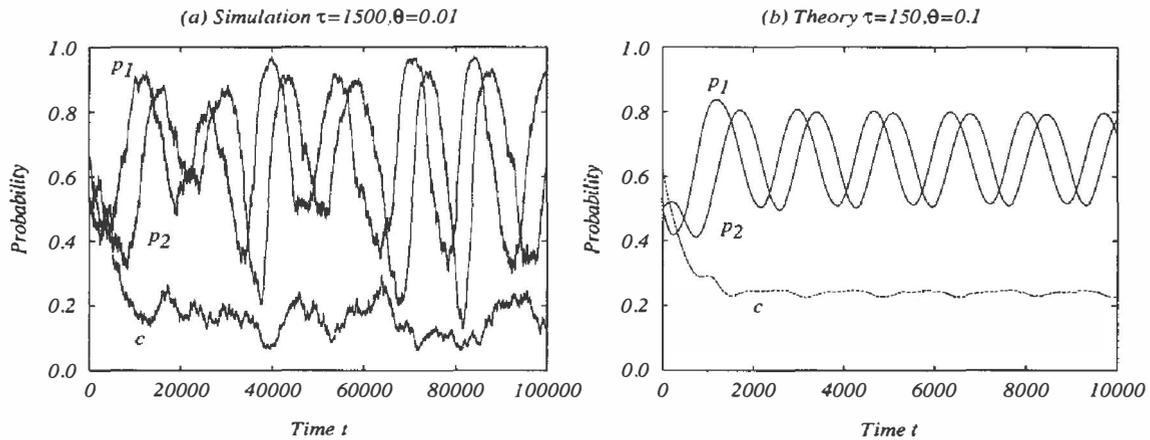

Figure 5: Oscillations in Action Strategies for Game 2

Figure 3 also shows that a large region of initialization is expected to result in the non-coalition equilibrium. For example, initialization $p_3=p_4=0.7$ is in the upper-right corner but is actually slightly biased to the non-coalition equilibrium ($c=p_3p_4=0.49$).

Figure 4 shows that players (theoretically) are able to reach mixed strategy equilibria (in both the group and action levels) for games 2 and 3 without delays. (Unless otherwise stated, $\alpha=0.02, \beta=0.4, \theta=0.01$ for game 2 and $\alpha=0.01, \beta=0.05, \theta=0.1$ for game 3). A distinction between the two games is that the likelihood of group formation, as defined by the clustering parameter $c$, decreases in game 2 and increases in game 3. We now consider the effects of delays in the information exchanged in these two games.

Figure 5 shows both a single simulation run and the prediction of (5) for game 2. In experiments with instabilities, it is not typically possible to combine multiple runs [Kephart et al., 1989]. Each particular run may grossly approximate theory, for example, by displaying persistent oscillations of appropriate amplitude and frequency. However, there are small phase shifts in the oscillations among multiple runs that lead to eventual obliteration of the oscillations after a long time. Instead, correlation must be attempted within each run and then averaged over multiple runs. We do not attempt to prove the correlation here but concentrate on the predictions of the theory. Independent of the accuracy of the dynamical equation with respect to the learning automata experiment, we consider the equation to be a paradigm for incremental learning.

We note that the accuracy is affected by the learning rate $\theta$: the smaller the learning rate, the better the accuracy. Large step sizes allow the strategies to overshoot the maxima and minima predicted. In Figure 5(a), the amplitudes in the simulation are larger than predicted but would be reduced if a smaller parameter value was chosen. In both cases, the relative delay is the same, i.e. $\theta\tau = 15$, although the individual parameters in the two cases differ by an order of magnitude. For this reason, we may examine the theory with any value of $\theta$, though we know that a small $\theta$ must be chosen to get an accurate simulation. Note that persistent oscillations are predicted (for the action strategies) and we can say that the delay to initiate such oscillations, $\tau_2$, must be less than or equal to 150 (for $\theta=0.1$). At lower values of delay, the theoretical strategies exhibit damped oscillations, however, simulations do not typically show the theoretical damping but rather noise in the strategies.

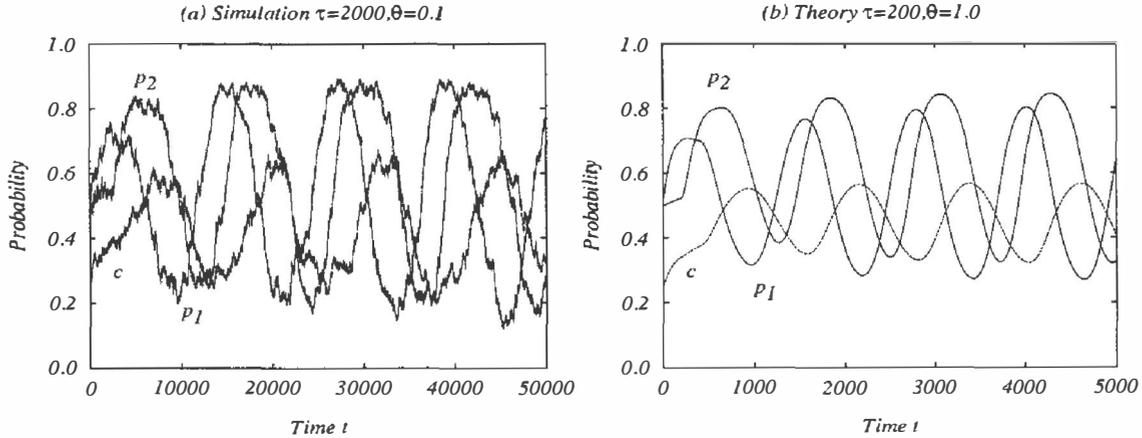

Figure 6: Oscillations in Action and Group Strategies for Game 3

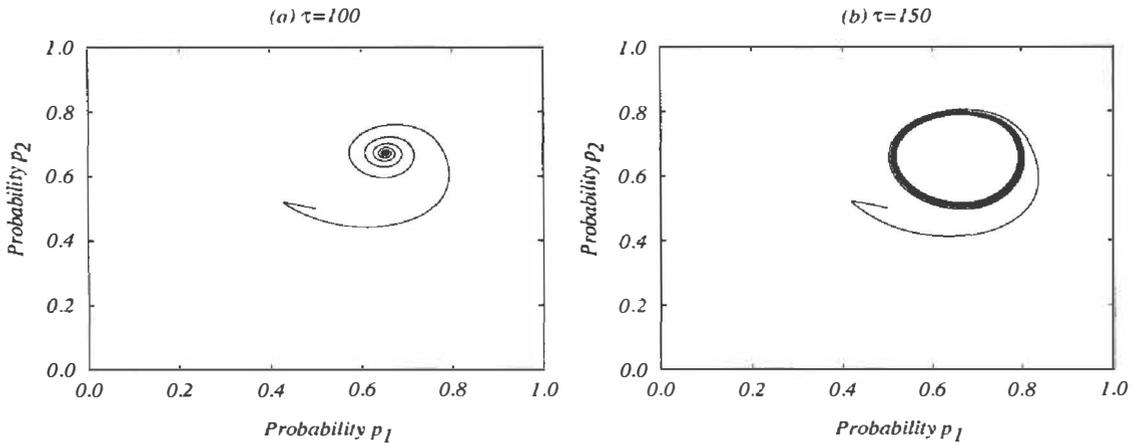

Figure 7: Phase-Plane Portraits of Action Strategies for Game 2 Near Stability Boundary

Figure 5 shows that the players reach a rough equilibrium in the group strategies for game 2 but Figure 6, for game 3, shows that the group strategies oscillate persistently. In this case, we can say that $\tau_2 <= 200$ for $\theta = 1.0$. (Other experiments with this game suggest that both the action and group strategies initiate oscillatory behavior at the same delay.) There is a rough approximation between theory and simulation, again with slightly higher amplitudes in simulation due to the step size parameter.

Figure 7 shows the predicted behavior of the action strategies for game 2 for two delays near the stability boundary between damped and persistent oscillations. The damped oscillations reach an equilibrium such that the center of the spiral vanishes and the same equilibrium serves as an attractor in the persistent oscillation case (i.e. limit cycle). Note that the circular nature of the phase-plane portrait in Figure 7(b) is an alternative display of the persistent oscillations, shown over time, in Figure 6(b). From Figure 7(a) and (b), we can conclude that $100 < \tau_2 \leq 150$.

Figure 8 shows the onset of a chaotic attractor, with corresponding shifting behaviors, at very high delay. As noted in Section 3, there is a complex interaction between the two levels of learning: action strategies affect group strategies and vice versa. The high delay in the experiment induces the strategies to revisit a variety of potential equilibria, but with small shifts in the trajectory. Figure 8(a) shows the specific behavior of the action strategies and Figure 8(b) shows the behavior of the group strategies. Together, these two figures demonstrate, in four-dimensional space, the complex dynamics of learning at two levels under the circumstance of delayed information.

## 5   Analysis

In this section, an approximation is used to determine the amount of delay $\tau_2$ required to initiate persistent oscillations. The technique involves linearizing in the neighborhood of the equilibrium $\mathbf{p}^*$ (a common approach [Kephart et al., 1989; Farmer, 1982]) and the assumption that $p_3$ and $p_4$ are constant and equal to the equilibrium values in $\mathbf{p}^*$, that is, $c = c^*$. This implies that we ignore the partial derivatives with respect to these variables. The resultant equations are

$$\frac{d\delta p_1}{dt} = X_1 \delta p_1 + Y_1 \delta p_2^\tau \quad (7)$$

$$\frac{d\delta p_2}{dt} = X_2 \delta p_2 + Y_2 \delta p_1^\tau \quad (8)$$

where

$$X_1 = \theta \frac{\partial W_1(\mathbf{p}^*)}{\partial p_1}, \quad X_2 = \theta \frac{\partial W_2(\mathbf{p}^*)}{\partial p_2},$$

$$Y_1 = \theta \frac{\partial W_1(\mathbf{p}^*)}{\partial p_2^\tau}, \quad Y_2 = \theta \frac{\partial W_2(\mathbf{p}^*)}{\partial p_1^\tau}.$$

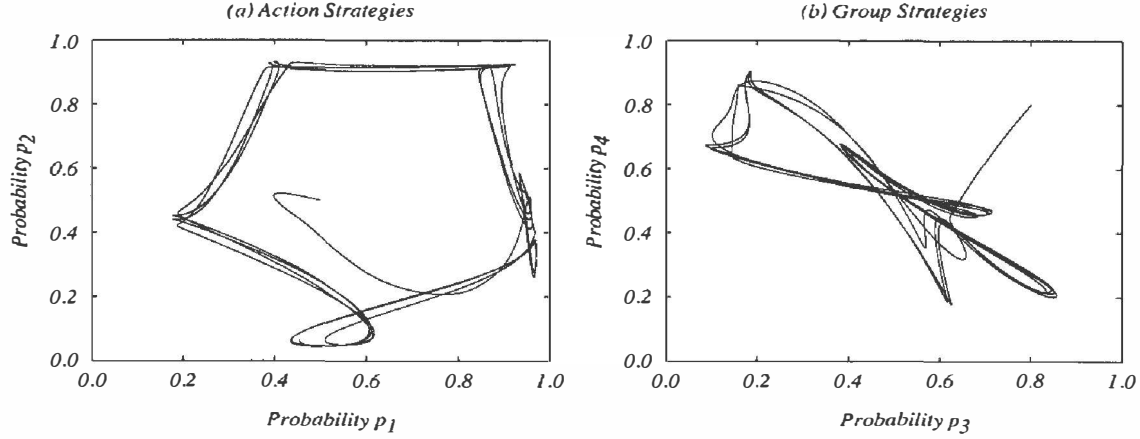

Figure 8: Chaotic Regime for Game 2 at High Delay ($\tau = 1000$)

The partial derivatives are straight-forward (see [Lakshmivarahan and Narendra, 1982] for a non-delay zero-sum version).

Assuming an exponential solution of the form $\delta p_1(t) = A_1 e^{\lambda t}$, $\delta p_1^\tau(t) = A_1 e^{\lambda(t-\tau)}$, etc. yields

$$(\lambda - X_1)(\lambda - X_2) = Y_1 Y_2 e^{-2\lambda\tau}. \quad (9)$$

Let $\lambda = r + iw$. There are an infinite number of discrete solutions and those parameter settings that yield only negative real parts are stable (with perhaps damped, but not persistent, oscillations). That is, marginal stability occurs at $r = 0$. The stability boundary can be determined by substituting $\lambda = iw$ in (9), applying Euler's formula, and solving for the real and imaginary parts:

$$\cos(2w\tau) = (X - w^2)/Y, \quad (10)$$
$$\sin(2w\tau) = (X_1 + X_2)w/Y, \quad (11)$$

respectively, where $X = X_1 X_2$ and $Y = Y_1 Y_2$.

Dividing (11) by (10),

$$\tan(2w\tau) = \chi = \frac{(X_1 + X_2)w}{X - w^2} \quad (12)$$

and the instability delay, sufficient to initiate persistent oscillations, is:

$$\tau = \tau_2 = \tan^{-1}(\chi)/2w, \quad (13)$$

where the inverse tangent takes its value in the interval $[0, \pi/2]$.

Adding the squares of (10) and (11),

$$u^2 + Bu + C = 0, \quad (14)$$

where

$$u = w^2, \; B = X_1^2 + X_2^2, \; C = X^2 + Y^2,$$

hence $w = \pm\sqrt{u}$. The single solution to the quadratic equation is

$$u = \frac{-B + \sqrt{B^2 - 4C}}{2}, \quad (15)$$

as the other solution fails to insure a real (the only type of solution) for $w$ (note that $B > 0$).

We are now in a position to predict the stability boundary between damped and persistent oscillations, the results are shown in Table 1 with the cases from Figures 5 and 6 included. The predicted values $\tau_p$ are based on (13). The observed values $\tau_o$ are not from simulation but from long runs of (5) at incremental delay to determine which delay is sufficient to initiate persistent oscillations to within a high degree of accuracy. There is close agreement between the values and we can draw three simple conclusions: $\tau_2$ increases with increasing $\alpha$, decreasing $\beta$, and decreasing $\theta$. The first two involve the relative strengths of the penalty and reward parameters. The adjustment of parameters to avoid instabilities under delayed information is exactly *opposite* the adjustments required to insure equilibria close to the optimal value of the game in a non-delayed environment. The last case is obvious from the fact that $\theta\tau$ is a measure of the relative delay; in fact, the table shows that the delay $\tau$ is doubled as the step size $\theta$ is halved. Finally, the data suggests that ignoring the partial derivatives with respect to $p_3$ and $p_4$ did not hinder the analytic prediction (even though these probabilities oscillated in game 3).

Table 1: Instability Delay: Predicted ($\tau_p$) versus Observed ($\tau_o$)

| game | $\alpha$ | $\beta$ | $\theta$ | $c^*$ | $\tau_o$ | $\tau_p$ |
|---|---|---|---|---|---|---|
| 2 | 0.02 | 0.80 | 0.1 | 0.2374 | 33 | 34 |
| 2 | 0.02 | 0.40 | 0.1 | 0.2417 | 145 | 148 |
| 3 | 0.01 | 0.10 | 1.0 | 0.6564 | 18 | 22 |
| 3 | 0.02 | 0.10 | 1.0 | 0.4812 | 52 | 51 |
| 3 | 0.01 | 0.05 | 1.0 | 0.4806 | 106 | 102 |
| 3 | 0.01 | 0.05 | 0.5 | 0.4793 | 218 | 203 |

## 6 Conclusions

A model has been presented with uncertainty in actions, group dynamics, payoffs, and state information. Learning automata achieve equilibrium in the particular cases examined with instantaneous information. This means that an agent successfully employs an automaton at each of the two levels. However, with delays in the system, the behaviors may exhibit damped or persistent oscillations and the onset of chaotic regimes.

The analysis yields the delay required to initiate persistent oscillations; unfortunately, the parameter settings that decrease the likelihood of instabilities also increase the likelihood that a suboptimal equilibrium will result. This illustrates the fundamental problem of seeking the optimum strategy without being misled by delayed information. However, the analysis is useful in that agents which communicate often enough to insure $\tau < \tau_2$ are guaranteed that persistent oscillations will not develop, thus insuring the stability of the system. This can have a strong impact on the performance of the system as stability is usually a prerequisite for good performance. In general, stability also is a measure of successful learning.